%% file: sumicoexplanation.tex
\documentclass[review,number,sort&compress]{elsarticle}
\usepackage{lineno}
\linenumbers
\usepackage{amsmath,amssymb}
\usepackage{enumerate}
\usepackage{graphics}
\journal{Nuclear Instruments and Methods in Physics Research A}

\begin{document}
\begin{frontmatter}

 \title{The Tokyo Axion Helioscope}
 \author[UTOKYO]{R. Ohta\corref{cor1}}
 \ead{comic@icepp.s.u-tokyo.ac.jp}
 \cortext[cor1]{Corresponding author.}
 \author[RIGAKUBU]{Y. Akimoto}
 \author[ICEPP,RESCUE]{Y. Inoue}
 \author[UTOKYO,RESCUE]{M. Minowa}
 \author[UKYOTO]{T. Mizumoto}
 \author[ICRR]{S. Moriyama}
 \author[ICEPP]{T. Namba}
 \author[TAKASU]{Y. Takasu}
 \author[KEK,RESCUE]{A. Yamamoto}

 \address[UTOKYO]{Department of Physics, School of Science, University of Tokyo, 7-3-1 Hongo, Bunkyo-ku, Tokyo 113-0033, Japan}
 \address[RIGAKUBU]{School of Science, University of Tokyo, 7-3-1 Hongo, Bunkyo-ku, Tokyo 113-0033, Japan}
 \address[ICEPP]{International Center for Elementary Particle Physics, University of Tokyo, 7-3-1 Hongo, Bunkyo-ku, Tokyo 113-0033, Japan}
 \address[UKYOTO]{Department of Physics, Graduate School of Sciene, Kyoto University, Kitashirakawa-oiwake-cho, Sakyo-ku, Kyoto 606-8502, Japan} 
 \address[ICRR]{Institute for Cosmic Ray Research, University of Tokyo, 456 Higashi-Mozumi, Kamioka-cho, Hida, Gifu 506-1205, Japan}
 \address[TAKASU]{Tsukuba, Ibaraki, Japan}
 \address[KEK]{High Energy Accelerator Research Organization, 1-1 Oho, Tsukuba, Ibaraki 305-0801, Japan}
 \address[RESCUE]{Research Center for the Early Universe (RESCEU), School of Science, University of Tokyo, 7-3-1 Hongo, Bunkyo-ku, Tokyo 113-0033, Japan}
 \begin{abstract}
  The Tokyo Axion Helioscope experiment aims to detect axions which are produced in the solar core.
  The helioscope uses a strong magnetic field in order to
  convert axions into X-ray photons and has a mounting to follow the sun very accurately. 
  The photons are detected by an X-ray detector which is made of
  16 PIN-photodiodes.
  In addition, a gas container and a gas regulation system is adopted for
  recovering the coherence between axions and photons in the conversion region giving sensitivity to axions with masses up to 2\,eV.
  In this paper, we report on the technical detail of the Tokyo Axion Helioscope.
 \end{abstract}
 \begin{keyword}
  solar axion \sep helioscope \sep PIN photodiode \sep superconducting magnet
 \end{keyword}
\end{frontmatter}
 \input{Introduction}

 \input{Tracking}
 \input{Magnet}
 \input{Gas_regulation_system}
 \input{PIN}
 \input{Summary}
 \input{Acknowledgement}

\end{document}

%% file: Introduction.tex
\section{Introduction}
\begin{figure}[hbt]
\begin{center}
  \resizebox{110mm}{!}{\includegraphics{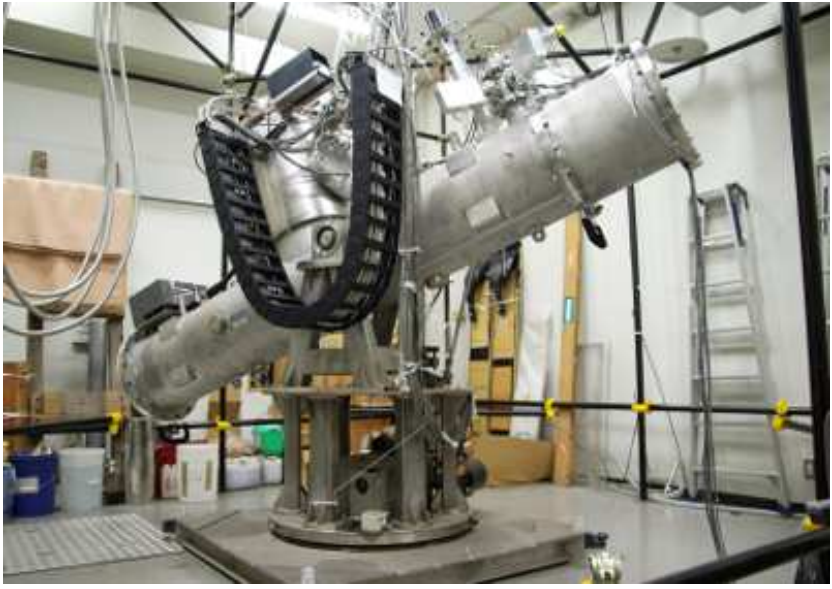}} 
  \caption{Photograph of Tokyo Axion Helioscope}
  \label{fig:photosumico} 
\end{center}
\end{figure}

Axions are light neutral pseudoscalar particles which are introduced to solve the strong CP problem~\cite{PQ}. 
They would be produced in the solar core via the Primakoff conversion of the plasma photons if they have large enough coupling to photons~\cite{Sikivie}~\cite{Bibber}.
The Tokyo Axion Helioscope experiment (also known as Sumico experiment) aims to detect solar axions by tracking the sun (Fig.~\ref{fig:photosumico}).
The helioscope consists of a tracking system, a
superconducting magnet, a gas regulation system, and an X-ray detector.
The magnet, the gas container and the X-ray detector are assembled in
a 3-m long cylinder.
The tracking system supports and drives the cylinder of the
helioscope to align the helioscope axis with the direction of the sun. 

Axions are converted into photons with energy of several keV, 
reflecting temperature of the solar core. 
These photons are detected by an X-ray detector.
The detection principle is based on the coupling of an incoming axion to a virtual photon provided by the transverse magnetic field of an intense dipole magnet.
 Thus the incoming axion is converted into a real, detectable photon that carries the energy of the original axion.
 The conversion rate is given by~\cite{Bibber}
\begin{equation}
  P_{a\rightarrow\gamma}=\left|\frac{g_{a\gamma\gamma}}{2}\exp\!\left[-\int_0^L \!\!\!\mathrm{d}z \ \Gamma/2\right]\right.\times \left.\int_0^L \!\!\!\mathrm{d}z B_\perp\exp\!\left[\mathrm{i}\!\int_0^z \!\!\!\mathrm{d}z^\prime\!\left(\!q\!-\!\frac{\mathrm{i}\Gamma}{2}\right)\right]\right|^2\!\!, \label{eq:prob}
\end{equation}
where $g_{a\gamma\gamma}$ is the coupling constant between axions and photons, $z$ and $z^\prime$ are the coordinate along the incoming axion,
$B_\perp$ is the strength of the transverse magnetic field, $L$ is the
length of the field along $z$-axis, $\Gamma$ is the X-ray absorption
coefficient of medium with which the conversion region is filled, $q=(m_\gamma^2-m_a^2)/2E$ is the momentum transfer by the virtual photon, and $m_\gamma$ is the effective mass of the photon which equals to zero in vacuum. 

Axions have not been discovered yet in any of the past experiments.
In the first phase~\cite{Moriyama} of our experiment, the conversion region was vacuum and the experimental limit is given to be $g_{a\gamma\gamma}<6.0\times 10^{-10} \, \mathrm{GeV^{-1}}$ (95\% C.L.) for $m_a<0.03 \, \mathrm{eV}$.
To detect heavy axions, the momentum transfer should be
kept negligible by introducing a dispersion matching medium into the conversion region.
For this purpose, helium gas was used in the second phase~\cite{Berota} and the limit is given to be $g_{a\gamma\gamma}<6.3-10.5\times 10^{-10} \, \mathrm{GeV^{-1}}$ (95\% C.L.) for $m_a<0.27 \, \mathrm{eV}$. 
The third phase~\cite{Akichan} was also performed using denser helium gas than that of the second one and the limit is given to be $g_{a\gamma\gamma}<5.6-13.4\times 10^{-10} \, \mathrm{GeV^{-1}}$ (95\% C.L.) for $0.84< m_a < 1.00 \, \mathrm{eV}$. 

In this paper, we report on the technical detail of the Tokyo Axion Helioscope.

%% file: Tracking.tex
\section{Tracking System}
\begin{figure}[hbt]
  \begin{center}
    \resizebox{110mm}{!}{\includegraphics{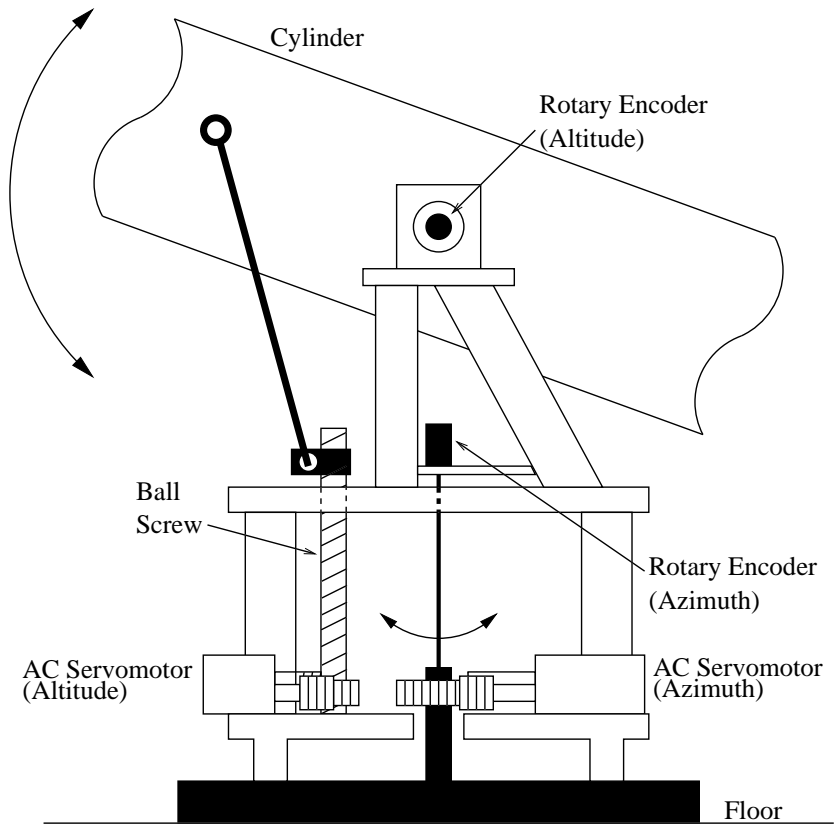}} 
    \caption{Schematic view of driving system}
    \label{fig:schemtrack} 
  \end{center}
\end{figure}

The tracking system consists of a driving mechanism, direction measuring encoders
and a PC (computer~1). They form a feedback control loop as a whole.
A schematic view of the driving system is shown in Fig.~\ref{fig:schemtrack}.
We adopted an altazimuth mount as the driving mechanism. The altazimuth mount minimizes inclination of the refrigerator and can support the magnet with simple structure.
The driving mechanism is composed of an axle bearing, a turntable, a ball screw, and two AC
servomotors.
The cylinder is supported by the axle bearing on the turntable. In the azimuthal direction, the turntable is driven by one of the servomotors (Hitachi EPS8).
In the altitudinal direction, the cylinder is driven by the other servomotor (Hitachi EPL3TD) through the ball screw.

The axion helioscope is located in Tokyo at $139^\circ45'48"\mathrm{E}$ and $35^\circ42'49"\mathrm{N}$.
In usual operation, the driving range of the helioscope is from $-28^\circ$ to $28^\circ$ in the altitudinal direction and the range of the azimuthal direction is $360^\circ$. 
Its azimuthal direction is limited only by a limiter which prevents the helioscope from endless rotation.
Within this limitations, the helioscope can track the Sun half of a day averagely.

\begin{figure}[hbt]
  \begin{center}
    \resizebox{110mm}{!}{\includegraphics{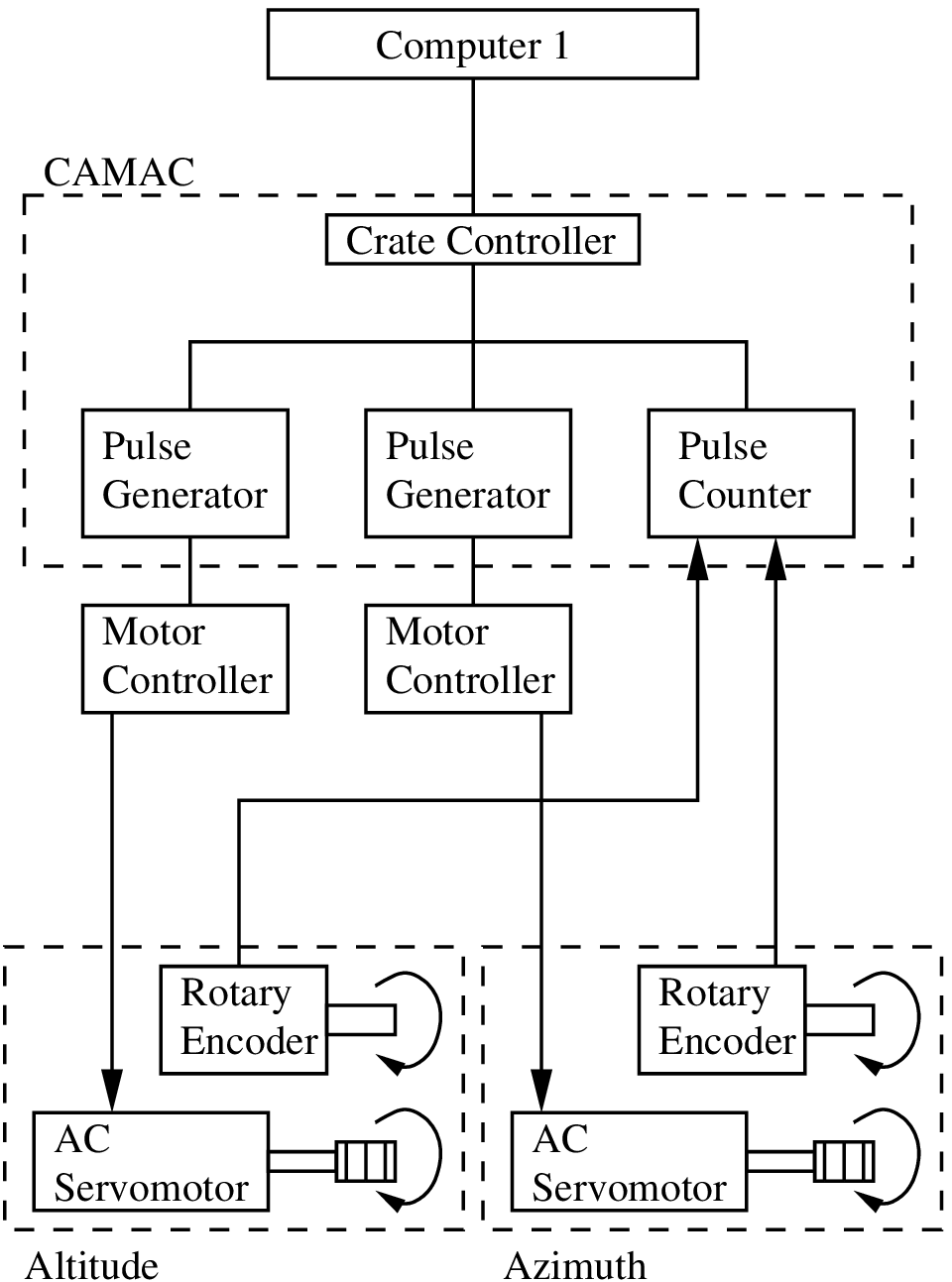}} 
    \caption{Control diagram of tracking system}
    \label{fig:diatrack} 
  \end{center}
\end{figure}

The control diagram of the tracking system is shown in Fig.~\ref{fig:diatrack}.
The driving mechanism is wholly controlled by computer~1 via CAMAC bus. 

The azimuthal and altitudinal directions of the helioscope are
measured by two precision rotary encoders (Canon R-1 L)
each of which produces 81,000 pulse-per-revolution quadrature
outputs in A and B channels and a zero index output in Z channel
to indicate its angular origin.
The encoder pulses are sent to a homemade pulse counter
by which computer~1 obtains the absolute angle of each encoders
with a resolution of
1/324,000 revolution or 4 arcseconds (ca. 19$\,\mu\rm rad$).

Each AC servomotor is controlled through a homemade CAMAC pulse generator. 
Every 0.2 second,
computer~1 calculates the amount and the frequency of the pulses,
namely the angle and the speed of rotation, for the next period
based on the measured- and targeted directions of the helioscope.

The overall tracking accuracy is better than 0.5\,mrad both in altitudinal and azimuthal direction. 
Main components of the errors are the fluctuation of the turntable and the misalignment of the magnet aperture and the helioscope axis.

The guidance of the helioscope movement is provided by the
tracking software.
In order to calculate the position of the sun, the U. S. Naval Observatory Vector Astronomy Subroutines (NOVAS-C ver 2.0.1)~\cite{NOVAS} is used.
NOVAS-C calculates the topocentric position of the sun with less than 2 arcseconds ($=9.7 \, \mu\mathrm{rad}$) error which is negligible in our measurements.

%% file: Magnet.tex
\section{Superconducting Magnet}
Incoming solar axions are converted into photons by magnetic field of
the superconducting magnet.
The conversion efficiency of axions depends strongly on the strength and
length of the magnetic field as shown in Eq.(\ref{eq:prob}).

Two devices make it easy to swing the magnet.
Firstly, the magnet is made cryogen-free by using two Gifford-McMahon refrigerators (GM refrigerators).
Therefore, no liquid helium tube is needed. 
Secondly, a persistent current switch (PCS) is equipped and the magnet is free from thick current leads.

\subsection{Magnet and Cryogenics}
\begin{figure}[hbt]
 \begin{center}
  \resizebox{110mm}{!}{\includegraphics{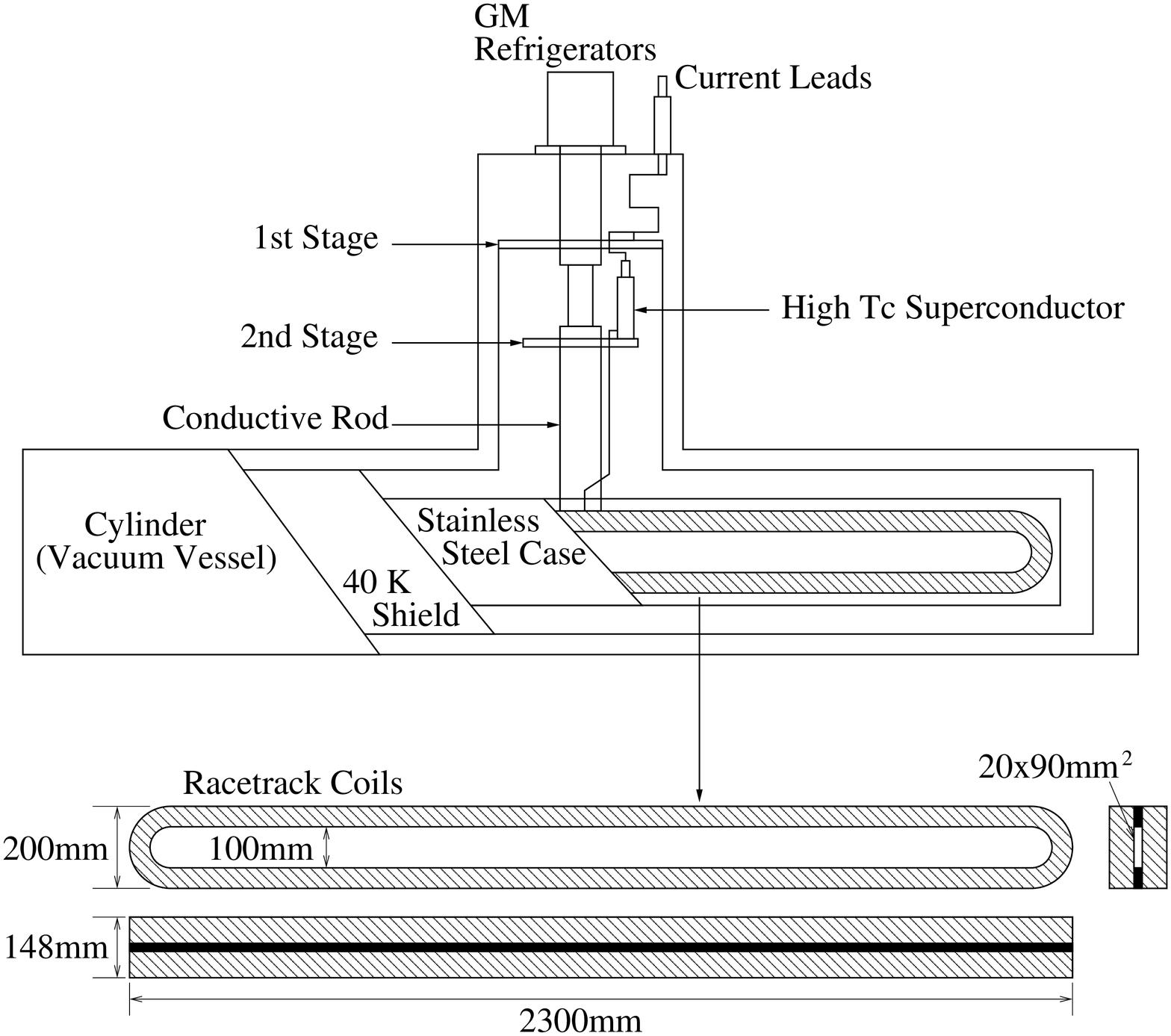}} 
  \caption{Schematic shape of race-track coils}
  \label{fig:schemagnet} 
 \end{center}
\end{figure}

The superconducting magnet consists of two 2.3-m long race-track shaped coils running parallel with a 20-mm wide gap between them. 
In this gap, the gas container is inserted and axions are converted into photons, which are to be detected by the X-ray detector.
A schematic view of the magnet is shown in Fig.~\ref{fig:schemagnet}. 

The coils are made of copper-clad NbTi superconducting wires and housed in a stainless steel case.  
They are kept at 5--6\,K during operation. 
The magnetic field in the gap is about 4\,T perpendicular to the helioscope axis when a current is 268\,A. 
The magnet is cooled by two GM refrigerators each of which is two staged.
The radiation shield of the cryostat is connected with the first-stage cold heads where the temperature reaches about 40\,K in normal operation. 
The coils are connected with the second stage, which reaches about 4\,K, through a thermal conductive rod and plates made of oxygen-free high-conductivity (OFHC) copper. High-$T_c$ superconducting current leads are used between the first and the second stages.

Thermometers are attached to the coils, cold heads of refregerators etc. to measure the temperature distribution. 
As thermometers, Carbon Glass Resistors (CGRs) and platinum-cobalt resistors (PtCos) are used.
CGRs are used to measure low temperatures (5--10\,K) and
PtCos are sensitive at relatively high temperatures (10--300\,K).
Each CGR was calibrated by the manufacturer.
Two Hall sensors are also attached to the coils to measure the strength
of the magnetic field.

\subsection{Persistent Current Switch}
\begin{figure}[hbt]
  \begin{center}
    \resizebox{110mm}{!}{\includegraphics{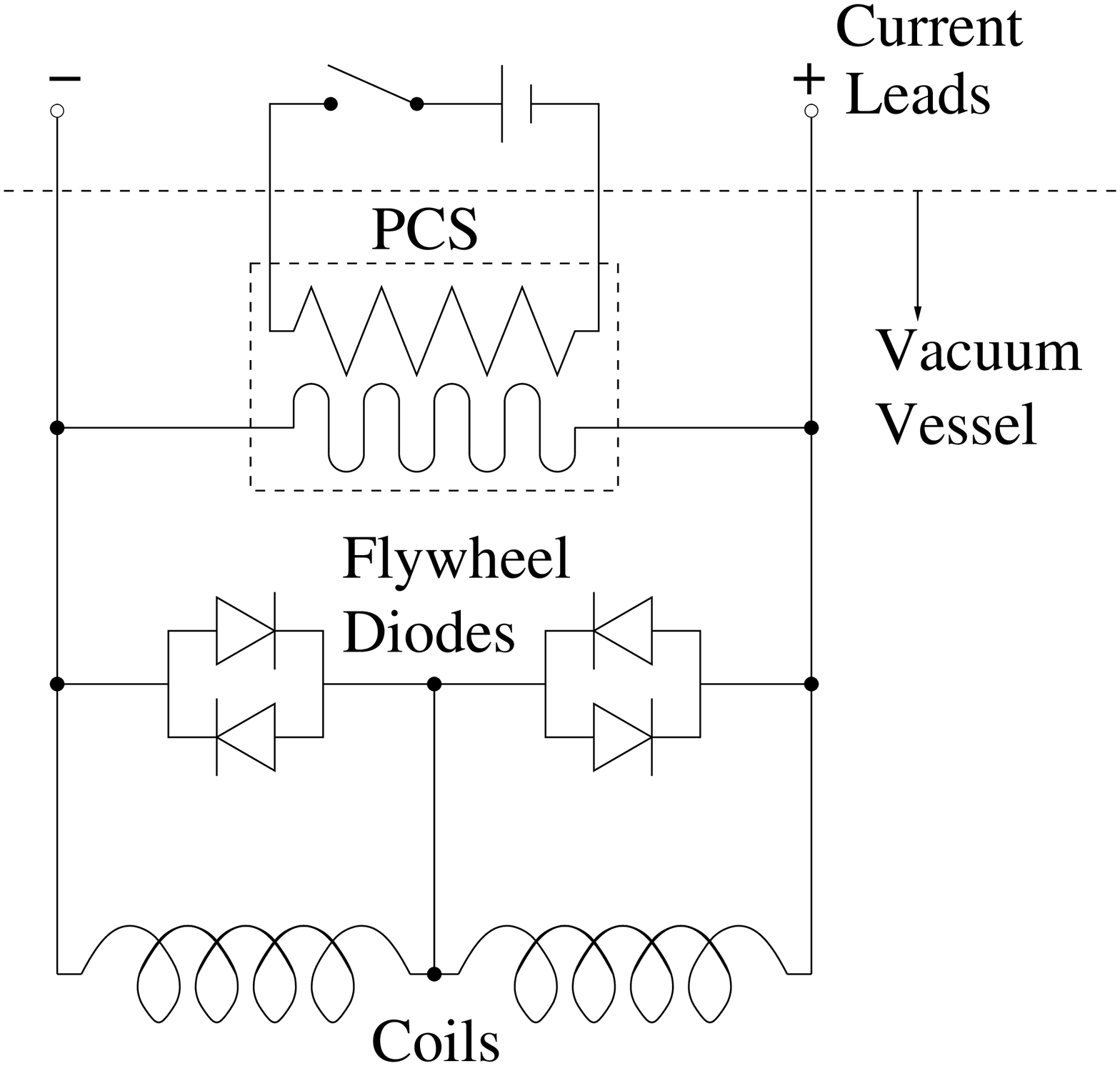}} 
    \caption{Circuit diagram of the magnet}
    \label{fig:diamagnet} 
  \end{center}
\end{figure}

The magnet can be operated in persistent current mode
by short-circuiting the coils with the persistent current switch
(PCS)~\cite{Mizumaki}.
In Fig.~\ref{fig:diamagnet}, a circuit diagram of the magnet is shown.
In this mode, the power supply is no longer needed to maintain
the magnetic field
and the current leads can be disconnected completely.

The PCS is mounted on a copper plate which is weakly coupled to the second
stage of the refrigerators.
The PCS is made of superconducting wire and an integrated heater to change
its phase to normal phase.
Before exciting the magnet, the heater is switched on and keeps the
temperature of the PCS above 9\,K, which is the phase-transition temperature
of the wire used in the PCS. The PCS in normal phase has resistance
around 11\,$\Omega$. 
Then a magnet excitation starts with supplying current.
When the current reaches the final value, excitation voltage goes to
zero and the heater is switched off to cool the PCS down.
After the PCS goes back to superconducting phase,
the supplying current can be decreased to zero while the current in the main circuit is maintained.

%% file: Gas_regulation_system.tex
\section{Gas Regulation System}
\input{Coherence}
\subsection{Gas Container}
\begin{figure}[hbt]
 \begin{center}
  \resizebox{130mm}{!}{\includegraphics{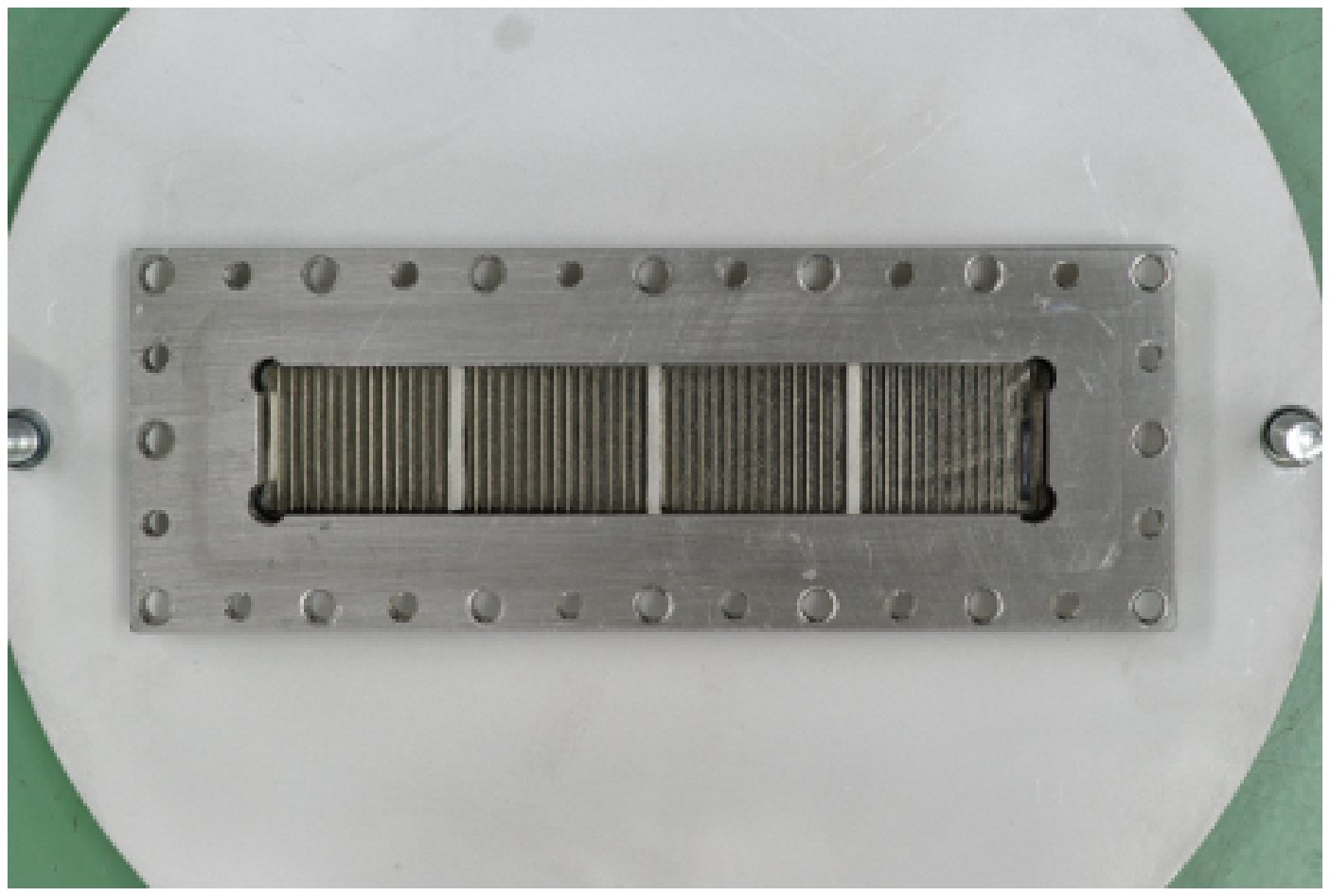}} 
  \caption{X-ray window}
  \label{fig:X-ray_window} 
 \end{center}
\end{figure}

The body of the container is made of four 2.3-m long 0.8-mm thick
stainless-steel rectangle pipes welded side by side to each other. 
The entire body is wrapped with 5N high purity aluminum sheet to achieve high uniformity of temperature. 
The measured thermal conductance between the both ends was 1 $\times$
10$^{-2}$\,W/K at 6\,K when the magnetic field is 4\,T. 
One end at the forward side of the container is sealed with welded plugs and is suspended firmly by three Kevlar cords, so that thermal flow through this end is highly suppressed. 
The opposite side nearer to the X-ray detectors is flanged and fixed to the magnet. 
At this end of the container, gas is separated from vacuum with an X-ray
window (Fig.~\ref{fig:X-ray_window}) manufactured by METOREX which is transparent to X-ray above 2\,keV
and can hold gas up to 0.3\,MPa at liquid helium temperature.
The window is composed of 25-$\mu$m thick Be foils, 1.5-$\mu$m Ni
support grids and Ni frames.
The foils have 1-$\mu$m polyimide coat on the one side.
Though He gas at 6\,K and 0.3\,MPa corresponds to the effective mass of about 4\,eV, absorption of the gas becomes dominant and sensitivity of the apparatus decreases in this mass region. 
\subsection{Pressure Control System}
\begin{figure}[hbt]
 \begin{center}
  \resizebox{110mm}{!}{\includegraphics{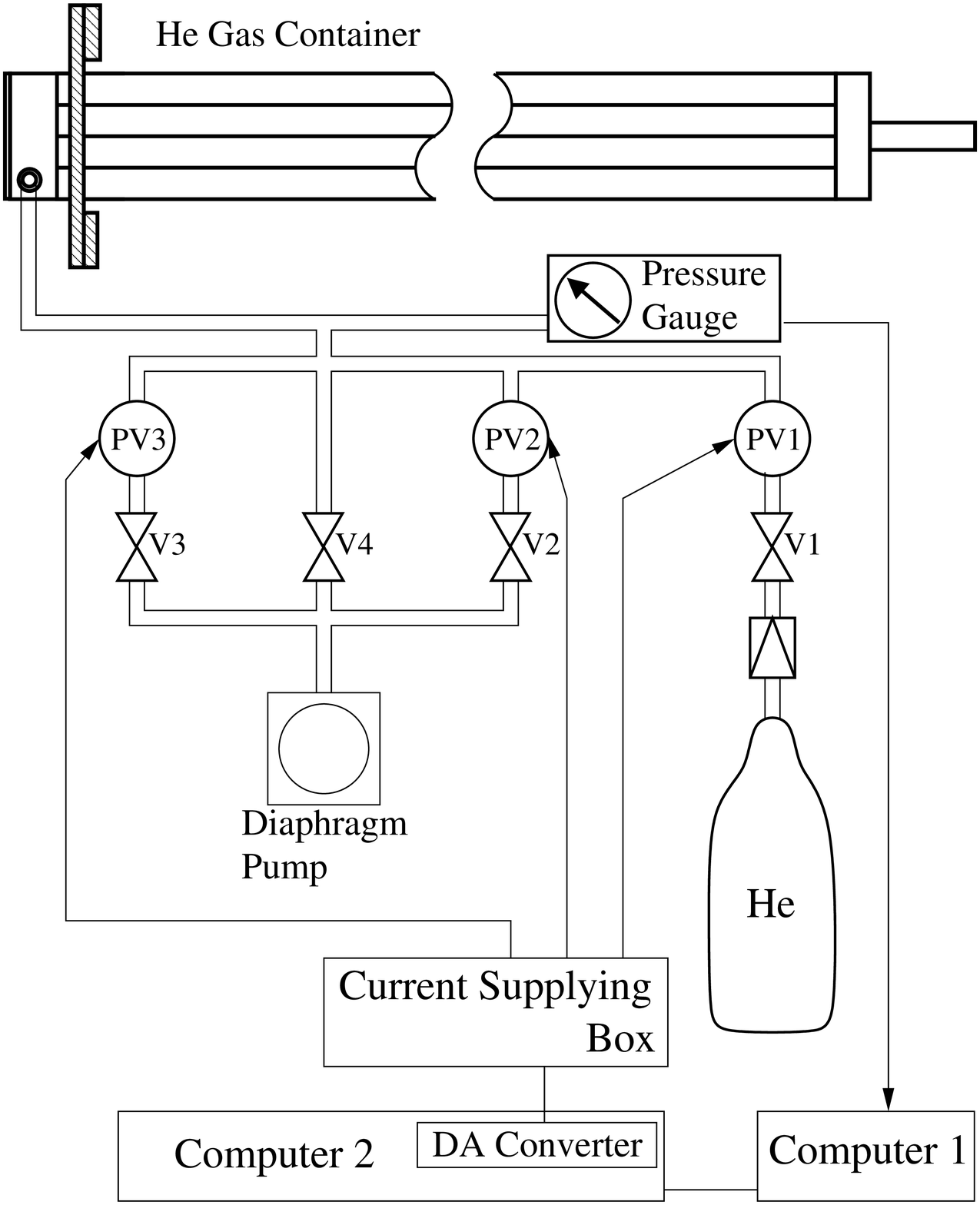}} 
  \caption{Diagram of pressure control system}
  \label{fig:schempres} 
 \end{center}
\end{figure}

To change effective mass of photon in the gas container, we developed the pressure control system of which diagram is shown in Fig.~\ref{fig:schempres}.
Main components of this system are the gas container, a precision pressure gauge (YOKOGAWA MU101-AH1N), a helium gas bottle for providing gas, a diaphragm pump (Pfeiffer Vacuum MVP 015-4) for suction, PCs for automatic regulation, and three piezo-valves (HORIBA STEC PV-1000 series).
The helium gas is injected through PV1 and is sucked by the vacuum pump through PV2 or PV3. 
The gas pressure is regulated by changing the gas flows through the piezo-valves which are controlled by their control voltages. 
Two PV-1101 piezo-valves are used in PV1 and PV2, while a PV-1302 is used in PV3.
PV3 is used to reach lower pressures below 10\,kPa, where the suction is not enough only with PV2. The gas flow of a PV-1302 is 30 times as high as a PV-1101.
Control voltages are supplied by computer~2 through a digital to
analog converter (Interface PCI-3521). 

In the third phase of this experiment, mass scanning was performed 
changing the effective mass by decreasing the pressure step by step. 
In fact, PV1 was opened only at the early stages of the scan until the gas pressure reached the highest value $p_s^{\rm max}$, and was closed during the automatic-control sequences.
The automated sequences worked as follows:
\begin{itemize}
 \item While the measured pressure $p_m$ was higher than the set pressure of a period $p_s$, PV2 (or PV3) was opened proportionally to $\Delta p = p_m - p_s$ as far as the flow rate was within the capability of the value. For larger $\Delta p$, PV2 (or PV3) was fully opened. The coefficient was determined empirically.
 \item Once $p_m$ became less than $p_s$, PV2 (or PV3) was
       closed and remained closed during measurement. 
 \item When a measurement had finished, $p_s$ was set to the next value and a new sequence was repeated.
\end{itemize}
 
\subsection{Temperature Control System}
Because the gas container is fixed on the magnet which is
thermally coupled to the GM refrigerators,
the temperature of the gas container is affected by fluctuations
of the cooling power of the refrigerators.
We developed a temperature control system,
since it is crucial to keep the gas temperature constant
to retain density uniformity of the gas as well as to control
its density reliably.

\begin{figure}[hbt]
 \begin{center}
  \resizebox{110mm}{!}{\includegraphics{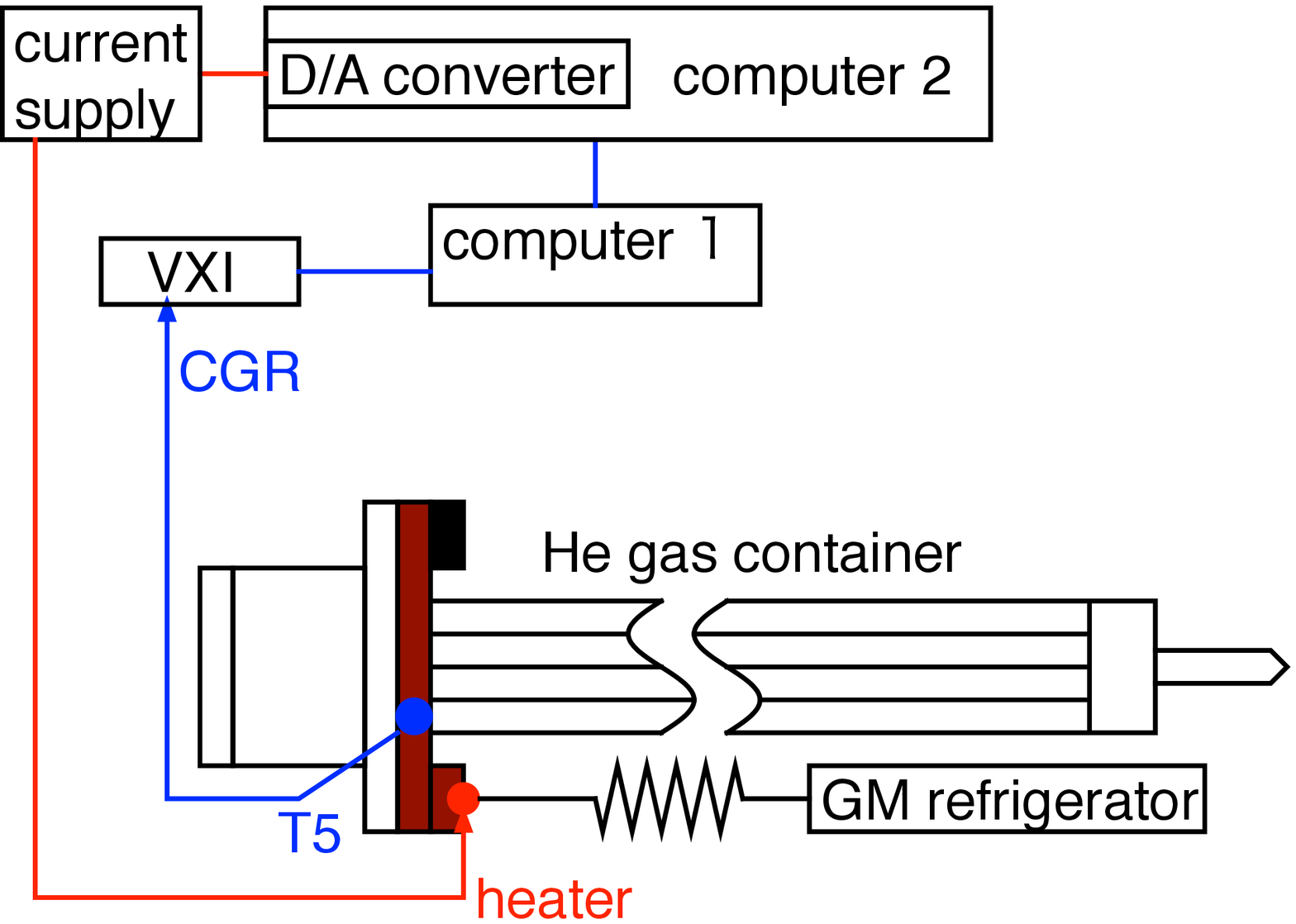}} 
  \caption{Diagram of temperature control system}
  \label{fig:temperature_control} 
 \end{center}
\end{figure}

The diagram of the temperature control system is shown in Fig.~\ref{fig:temperature_control}.
The main components of this system are a heater and a CGR thermometer T5.
T5 is measuring the temperature of a copper plate which is tightly
joined with the gas container.
The heater is attached to the heat link between the copper plate and
the magnet.
The current through the heater is controlled to keep the T5 temperature
constant, where PI (Proportional-Integral) control is adopted.
The PI parameters are decided empirically.

%% file: Coherence.tex
\subsection{Coherence between Axion and Photon}
To keep coherence between axion and photon, $m_\gamma$ has to be tuned to
$m_a$ by introducing gas into the conversion region.
For this purpose, the gas container is inserted into the aperture of the magnet.

A photon in the X-ray region acquires a positive effective mass in a medium. 
It is well approximated by
\begin{equation}
  m_\gamma = \sqrt{\frac{4\pi\alpha N_e}{m_e}} \label{eq:dens}
\end{equation}
as for the light gas, such as hydrogen or helium, where $\alpha$ is the fine structure constant, $m_e$ is  the electron mass, and $N_e$ is the number density of electrons~\cite{Bibber}.
Cold helium gas was adopted as a dispersion matching medium, because helium gas has small X-ray absorption length and remains at gas state at 5\,K, the operating temperature of our magnet, at 1\,atm.
If $m_a$ differs from $m_\gamma$, coherence between axion and photon is lost and the conversion efficiency decreases.
From Eq. (\ref{eq:prob}), the condition in which the conversion efficiency has significant value is $qL\lesssim\pi$.
For a given effective mass, axion masses $m_a$ within a very narrow range satisfy this condition.
The width of this window is about 2\,meV (FWHM)
at $m_\gamma\simeq 1 \, \mathrm{eV}$.
To sweep the axion mass range up to a few eV,
we need to change the gas density more than a thousand times.
To control the gas density, we need to control the pressure accurately and to keep the temperature constant.
Therefore we developed an automatic pressure and temperature control system of helium gas in the gas container for the third phase of solar axion search~\cite{Akichan}.

%% file: PIN.tex
\section{X-ray detector}
\subsection{PIN Photodiode X-ray Detector}
\begin{figure}[hbt]
 \begin{center}
  \resizebox{110mm}{!}{\includegraphics{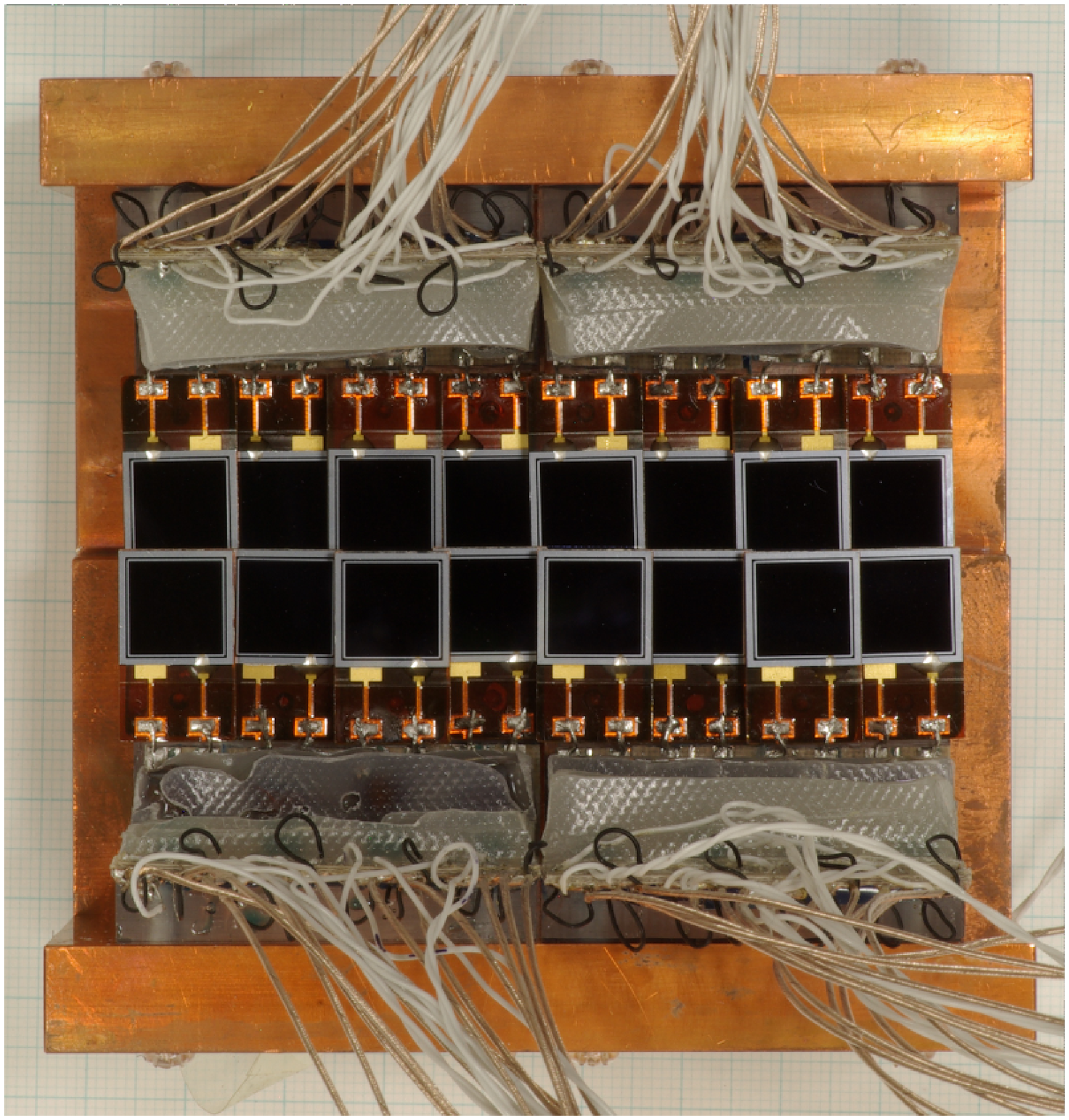}}
  \caption{Photograph of PIN photodiode array}
  \label{fig:photoPIN} 
 \end{center}
\end{figure}

The X-ray detector to detect photons originating from axions consists of 16 PIN photodiodes, Hamamatsu Photonics S3590-06-SPL(Fig.~\ref{fig:photoPIN}), and 16 preamplifiers which amplify signals from the PIN photodiodes. 

Each of preamplifiers is separated into two part called head and back-end part.
The head parts are mounted next to the photodiodes and include the first stage FETs, feedback resistors, test pulse inputs and HV buffers. The PIN photodiodes and the head parts are assembled in the helioscope cylinder. The remaining circuits are placed out of the cylinder.
The details of this X-ray detector is described elsewhere~\cite{NambaPIN}~\cite{AkiPIN}.
\subsection{Data Acquisition System}
\begin{figure}[hbt]
 \begin{center}
  \resizebox{110mm}{!}{\includegraphics{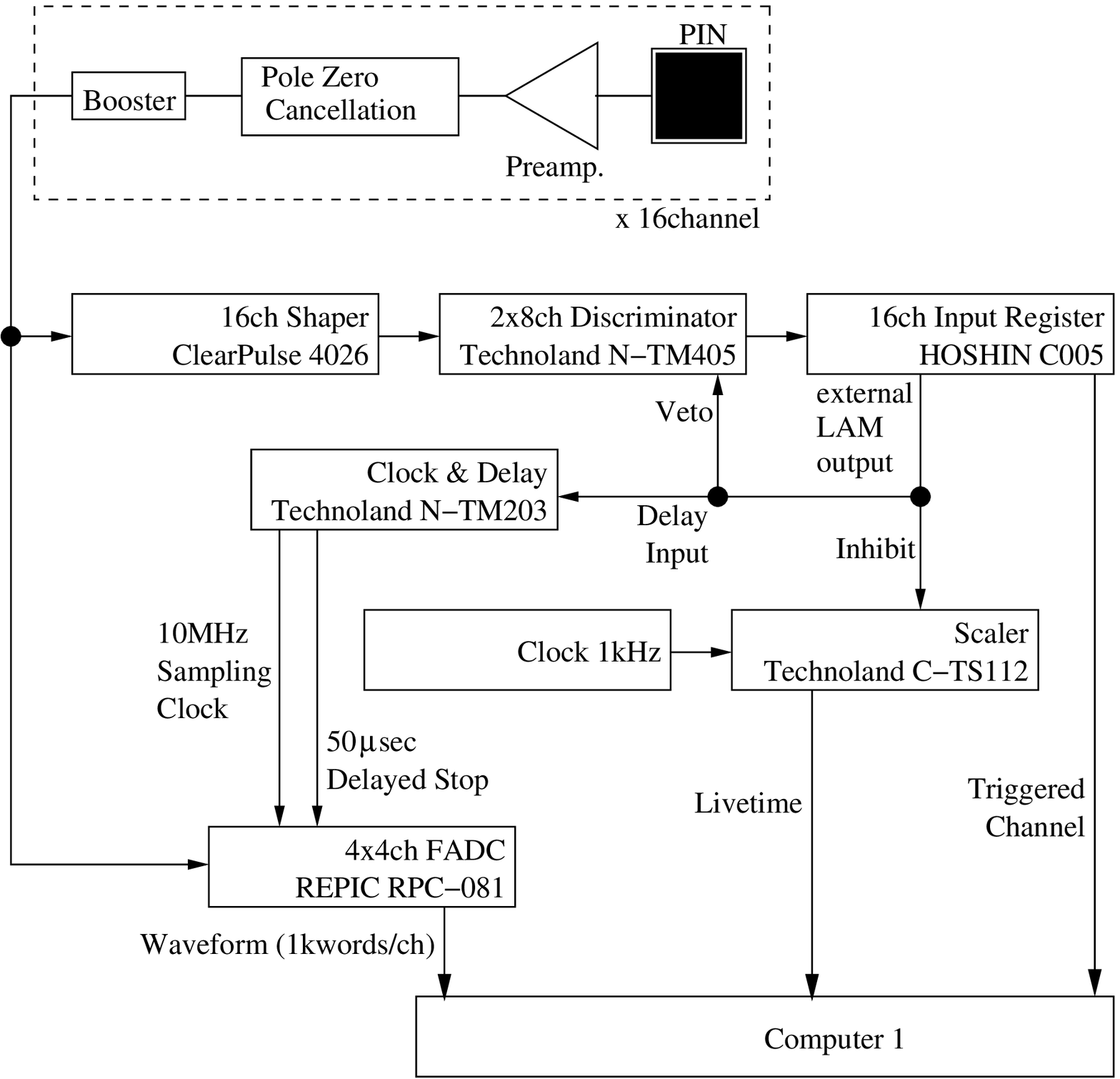}} 
  \caption{Diagram of DAQ system}
  \label{fig:DAQ} 
 \end{center}
\end{figure}

The output signals of the preamplifiers are stored and analyzed so that
we will obtain energy spectrum of the signals. The data acquisition system is shown in Fig.~\ref{fig:DAQ}.

Firstly, signals from preamplifiers are filtered and low frequency
components are reduced. Since the cut off frequency of this filter is 3 kHz, it does not affect X-ray signals whose rise time is 0.8--1.3\,$\mu\mathrm{s}$ but reduce microphonic noise. Next, the filtered signal is amplified by a linear amplifier called the booster with a gain of about 300 to fit the input range of the FADCs. These filter and booster are mounted just beside the back-end parts of the preamplifiers.

After the booster, signals are sent to the FADCs (REPIC RPC-081) via 15-m coaxial cables. The sampling period of the FADC is 0.1\,$\mu\mathrm{s}$, the resolution is 8 bit, and the buffer depth is 1024 words. 
The FADCs are stopped 50\,$\mu\mathrm{s}$ after the trigger signal is latched in the input registers (HOSHIN C005).
The trigger is made from a discriminator (Technoland N-TM405) and a shaping amplifier (Clear-Pulse 4026) whose shaping time is 3\,$\mu\mathrm{s}$. 
When the FADC is stopped by trigger, the input register tell the trigger to computer~1 and it takes waveform from the FADCs via CAMAC bus.
At the same time the livetime counter is stopped. 
Because deadtime of each transfer is about 0.35\,s and cannot be ignored
in the measurement, the livetime is counted in millisecond accuracy by a
1-kHz clock. 
This accuracy is needed to compare the solar tracking event and the non-tracking event.

%% file: Summary.tex
\section{Summary}
We developed the axion helioscope which aims to detect solar axions.
This experimental apparatus consists of the tracking system, the superconducting magnet, the gas regulation system, and the X-ray detector.
The tracking system directs the main cylinder towards the sun.
The magnet converts incoming axions into photons and the unmanned gas regulation system enables us to convert axions with masses up to eV scale.
The X-ray detector measures conversion photons and the DAQ system records the signal of the photons into the computer.
Using this apparatus, we have performed solar axion searches with axion's mass up to 0.27\,eV~\cite{Moriyama}~\cite{Berota} and around 1\,eV~\cite{Akichan}.
Since we did not detect an evident signal of axion,
we are going to perform more massive solar axion search with mass up to
about 2\,eV in the near future.

%% file: Acknowledgement.tex
\section{Acknowledgements}
The authors thank the former director general of KEK, Professor H. Sugawara, for his support in the beginning of the helioscope experiment. This research was partially supported by Grant-in-Aid for COE Research, the Japanese Ministry of Education, Science, Sports and Culture (MEXT) and Grant-in-Aid for Scientific Research (B), Japan Society for the Promotion of Science and also by the Matsuo Foundation. Additional support was provided by Global COE Program ``Physical Sciences Frontier'', MEXT, Japan.